\documentclass[a4paper]{iopart}
\usepackage{iopams}
\usepackage{amssymb,mathrsfs}
\usepackage{graphicx}
\usepackage[pdfauthor={Francesco Buscemi},pdftitle={Private quantum decoupling and secure disposal of information},pdftex,colorlinks=false]{hyperref}
\def\Tr{\mathrm{Tr}} \def\d{\mathrm{d}}
\def\>{\rangle}\def\<{\langle} \def\sH{\mathscr{H}}

 \def\povm#1{\mathbf{#1}}

\def\id{\mathrm{id}}
\DeclareRobustCommand\openone{\leavevmode\hbox{\small1\normalsize\kern-.33em1}}

\def\r{\varrho_{RA}} %\def\s{\varsigma^{A_1A_2B}}

\def\eps{\varepsilon}
\def\sn{\varsigma_{R_nB_nE_n}^{(n)}}
\def\s{\varsigma_{RBE}}
\def\mD{\mathcal{D}}

\newtheorem{corollary}{Corollary}

\newtheorem{prop}{Proposition}

\begin{document}

\title{Private Quantum Decoupling and Secure Disposal of Information}

\author{Francesco Buscemi}
\address{Institute for Advanced Research, Nagoya University, Japan}
\ead{buscemi@iar.nagoya-u.ac.jp}

\date{\today}

\begin{abstract}

Given a bipartite system, correlations between its subsystems can be understood as information that each one carries about the other. In order to give a model-independent description of secure information disposal, we propose the paradigm of \emph{private quantum decoupling}, corresponding to locally reducing correlations in a given bipartite quantum state without transferring them to the environment. In this framework, the concept of \emph{private local randomness} naturally arises as a resource, and total correlations get divided into eliminable and ineliminable ones. We prove upper and lower bounds on the amount of ineliminable correlations present in an arbitrary bipartite state, and show that, in tripartite pure states, ineliminable correlations satisfy a monogamy constraint, making apparent their quantum nature. A relation with entanglement theory is provided by showing that ineliminable correlations constitute an entanglement parameter. In the limit of infinitely many copies of the initial state provided, we compute the regularized ineliminable correlations to be measured by the coherent information, which is thus equipped with a new operational interpretation. In particular, our results imply that two subsystems can be privately decoupled if their joint state is separable.

\end{abstract}

%{\footnotesize \tableofcontents}

\maketitle

%%%%%%%%%%%%%%%%%%%%%%%%%%%%%%%%%%%%%%%%%%%%%%%%%%%%%%%%%%%%%%%%%%%%%%

\section{Introduction}\label{sec:1}

Let us suppose we are given a medium containing sensitive information which, for some reason, we want to dispose of in a secure way, i.~e. in such a way that neither we nor anyone else can have access to---or be deemed to possess---that information anymore. In many everyday situations, particular precautions have to be taken in advance in order to counter unwanted \emph{data remanence}, i.~e. the persistence of data that were nominally erased or removed.

When dealing with macroscopic objects, the irreversibility of dissipative processes is generally enough to provide a \emph{practically secure} erasure of data (think of e.~g. shredding the medium). However, such a conclusion is \emph{in-principle} completely inadequate. Let us suppose in fact that the medium carrying the information is represented by the state of a microscopic object obeying the laws of Quantum Mechanics. Since quantum evolutions are globally reversible, also information has to be globally preserved, and any sort of \emph{true} information erasure is thus forbidden~\cite{no-del}. Even so, information could be, if not erased, \emph{hidden} or encoded in such a way to achieve what a secure disposal of information is meant to achieve. Our aim here is to introduce and analyze a model-independent paradigm of secure information disposal, suitable to describe both classical and quantum information disposal, and naturally encompassing the ideal situation in which the only limitations to data processing are those imposed by the laws of Quantum Mechanics.

In order to do so, let us consider the following protocol: let $\varrho_{RA}$ denote the initial bipartite state shared between two parties---the active player (or \emph{receiver}) $A$ and the passive and inaccessible reference system (or \emph{remote sender}) $R$. According to a common understanding, the amount of correlations existing in $\varrho_{RA}$ between subsystems $A$ and $R$ can be interpreted as the amount of information that $A$ carries or possesses \emph{about} $R$. Since the content of the message is assumed to be private (otherwise there is no reason to require security in its disposal), we suppose that the state $\varrho_{RA}$ is decoupled from other accessible quantum systems, in particular from the local environment---i.~e. the `trash' system---of $A$. The goal for $A$ is to securely dispose of the information she has about $R$. By identifying information with correlations, this means that she has to reduce her correlations with $R$ by applying local operations on her share only (as $R$ is not accessible), \emph{and} without transferring any of these correlations into her local environment (as this is assumed to be accessible to adversaries). We name this protocol \emph{(local) private quantum decoupling} (PQD).

PQD constitutes a novel instance of the general task of producing, under various constraints, an uncorrelated (or, equivalently, `factorized') state out of a correlated one. The importance of studying decoupling protocols lies in the fact that, with appropriate constraints, it is possible to quantitatively characterize diverse properties of quantum correlations with respect to their robustness \emph{against} decoupling. Such an approach has been introduced independently in Refs.~\cite{groisman} and~\cite{malloppo}, and contributed in recognizing the central role that decoupling plays in quantum information processing: for example, the primitives of state merging~\cite{state-merg}, state redistribution~\cite{state-redistr}, and the `mother' protocol~\cite{mother}, are all based on decoupling arguments, and decoupling procedures form the building blocks of many recently constructed coding theorems achieving quantum capacity~\cite{coding,one-shot}. Hence, PQD offers a new point of view on the study of quantum correlations, with possible implications in entanglement theory and quantum cryptography.

The structure of the paper is as follows: in Section~\ref{sec:2}, previous approaches to quantum decoupling are described and the new paradigm of PQD is motivated through simple examples. The rigorous definition of PQD is given in Section~\ref{sec:3} by defining eliminable and ineliminable correlations, and the concept of private local randomness as a resource is introduced. In Section~\ref{sec:4} we prove general bounds on PQD when acting on an arbitrary mixed state, and show that ineliminable correlations are in fact monogomous correlations, in the sense that they cannot be freely shared. This is a feature common to many distinctively quantum correlations, like e.~g. entanglement (when suitably measured). Section~\ref{sec:5} deals with the asymptotic limit where infinitely many identical and independent copies of a given state are available: in this case, the optimal rate for PQD is explicitly calculated as being expressed by the coherent information. In Section~\ref{sec:6} we discuss some examples for which PQD assumes a particularly simple form. A connection between PQD and random-unitary channels is exhibited, together with an open question concerning the latter. Finally, Section~\ref{sec:7} describes the relations existing between ineliminable correlations, quantum entanglement, and other `quantum' correlations present in an arbitrary bipartite quantum state. Here we show that ineliminable correlations represent an entanglement parameter, in the sense given by~\cite{ent-param}. Section~\ref{sec:8} concludes the paper with a brief summary of the results obtained and possible directions to investigate in future.

\subsection{Notation and basic concepts}

All quantum systems considered in the following are finite dimensional, in the sense that their attached Hilbert spaces $\sH$ are finite dimensional. We use Greek letters like $\psi,\Psi,\cdots$ for pure quantum states, while letters like $\varrho,\varsigma,\tau,\cdots$ are reserved for mixed states. The usual ket-bra notation $|\psi\>\<\psi|$ denoting the rank-one projector onto the state $|\psi\>$ is generally abbreviated simply as $\psi$. Roman letters label the systems sharing a quantum state: for example, $\varrho_{RA}$ is a mixed state defined on the composite system $RA$, carrying the Hilbert space $\sH_R\otimes\sH_A$. Where no confusion arises (or it is not differently specified), omission of a letter in the label indicates a partial trace, namely, $\varrho_A:=\Tr_R[\varrho_{RA}]$.

In classical information theory, given two random variables $\mathbf{X}$ and $\mathbf{Y}$ distributed according to a joint probability distribution $p(x,y)$, the \emph{mutual information} $I(\mathbf{X}\!:\!\mathbf{Y}):=H(\mathbf{X})+H(\mathbf{Y})-H(\mathbf{XY})$, where $H(\mathbf{X}):=-\sum_xp(x)\log_2p(x)$ is the Shannon entropy, is known to measure both the degree of correlation existing between $\mathbf{X}$ and $\mathbf{Y}$ \footnote{Since $I(\povm{X}\!:\!\povm{Y})=0$ iff
  $p(x,y)=p(x)p(y)$ and $I(\povm{X}\!:\!\povm{Y})=H(\povm{X})$ iff
  $p(x,y)=p(y)\delta(x,f(y))$, namely, $\povm{X}$ is a function of
  $\povm{Y}$. See Ref.~\cite{cover-thomas}.}, as well as ``the reduction in the uncertainty of $\mathbf{X}$ due to the knowledge of $\mathbf{Y}$'' (see Ref.~\cite{cover-thomas} p.20). We hence say that $\mathbf{Y}$ carries a total of $I(\mathbf{X}\!:\!\mathbf{Y})$ bits of information about $\mathbf{X}$.

The classical arguments given above can be straightforwardly generalized to the quantum case as follows. According to Refs.~\cite{groisman,malloppo}, given a bipartite quantum state $\varrho_{RA}$, the \emph{quantum mutual information} (QMI)~\cite{qminfo} defined as $I^{R:A}(\varrho_{RA}):=S(\varrho_R)+S(\varrho_A)-S(\varrho_{RA})$, where $S(\sigma):=-\Tr\sigma\log_2\sigma$ is the von Neumann entropy of a state $\sigma$ and $\varrho_{R(A)}:=\Tr_{A(R)}[\varrho_{RA}]$, provides a sound and operationally meaningful measure of the total amount of correlations present in $\varrho_{RA}$ \footnote{This means that also correlations, as information, are measured here in (quantum) bits.}. It is known that $I^{R:A}(\varrho_{RA})=0$ if and only if $\varrho_{RA}=\varrho_R\otimes\varrho_A$, while $I^{R:A}(\varrho_{RA})=2S(\varrho_R)$ if and only if $\varrho_R$ is purified by $A$. Moreover, when systems $R$ and $A$ are classical, QMI coincides with its classical counterpart. It makes sense then to say that the system $A$ carries a total of $I^{R:A}(\varrho_{RA})$ (quantum) bits of information about $R$. Notice that the information content, as we defined it, turns out to be positive and symmetric, that is, the amount of information $A$ carries about $R$ equals the amount of information $R$ carries about $A$, since $I^{R:A}(\varrho_{RA})=I^{A:R}(\varrho_{RA})$. When the state $\varrho_{RA}$ is clear from the context, we will simply write $I(R\!:\!A)$ to indicate $I^{R:A}(\varrho_{RA})$, and, in case the state is multipartite like, for example, $\s$, we will denote $I^{R:B}(\Tr_E[\s])$ simply by $I^{R:B}(\s)$ or even by $I(R\!:\!B)$.

We finally recall here also the notion of \emph{coherent information}~\cite{cohinfo} which is well-known in the literature to play a fundamental role in entanglement theory and quantum Shannon theory and is defined as $I_c^{A\to R}(\varrho_{RA}):=S(\varrho_R)-S(\varrho_{RA})$.  Contrarily to QMI, coherent information can be negative and it is not symmetric as there is a preferred directionality in its definition.

\section{Quantum decoupling: previous definitions and new
  motivations}\label{sec:2}

Suppose two players, Alice ($A$) and a Referee ($R$), share a bipartite
quantum state $\r$. We say that they are \emph{decoupled} if and only
if $\r=\varrho_R\otimes\varrho_A$. If the initial state is not
decoupled, a decoupling procedure aims at transforming the initial
state $\r$ into (something close to) a factorized state. It is clear
that, without imposing any constraint on the decoupling procedure, the
whole task is trivial: one can always \emph{prepare}, even
\emph{locally}, an exactly decoupled state after having discharged the
input into the environment. If we want to probe some property of
quantum correlations using decoupling, the main point is then
to choose appropriate constraints making the best decoupling strategy
non-trivial. In the following, before presenting our decoupling task,
we will review two different types of decoupling tasks adopted in the
literature.

A first decoupling task is introduced through information-theoretical
arguments. Refs.~\cite{groisman,malloppo}, with analogous though
inequivalent arguments, prove that quantum mutual information
$I^{R:A}(\r)$ provides an \emph{operational} measure of total (both
classical and quantum) correlations shared between $A$ and $B$,
in that $I^{R:A}(\r)$ quantifies the cost of erasing initial correlations. In particular,
the decoupling protocol adopted in Ref.~\cite{groisman} is related
with the theory of \emph{private quantum channels}~\cite{private} and
aims at constructing a local map $A\to B$ of the form
\begin{equation}\label{eq:e-rand}
  \mD:\sigma\mapsto\frac 1N\sum_{i=1}^NU_i\sigma U_i^\dag,
\end{equation}
such that $(\id_R\otimes\mD_A)(\r)$ is close to some factorized state
$\varrho_R\otimes\omega_B$. The additional requirement posed in~\cite{groisman} is to use
in~(\ref{eq:e-rand}) the smallest possible number $N$ of unitary
operators (indeed, the average over the whole special unitary group
$\mathbb{SU}(d)$ achieves perfect decoupling, but it requires an
infinite amount of randomness to be added). Clearly, the number $N$
can in general depend both on the input state $\r$ and on the degree
of approximation required of the decoupling. Ref.~\cite{groisman}
proves that, in the i.i.d. asymptotic limit, exact decoupling can be
achieved for values of $N$ as small as $2^{nI(R:A)}$, where $n$ is the
(large) number of copies of $\r$ provided. Following Landauer's
principle, quantum mutual information is then interpreted as the total
amount of correlations \emph{per copy}, being the minimum cost (measured in
bits of extra-randomness borrowed from a randomness reservoir) needed
in the i.i.d. limit to achieve exact decoupling. A similar, though not
equivalent, conclusion can be drawn as a special case of the general
setting introduced in Ref.~\cite{malloppo}, where a class of
transformations different from~(\ref{eq:e-rand}) is considered (in
particular, classical communication between the two parties is taken
into account in an essential way). Ref.~\cite{malloppo}
interprets quantum mutual information as a particular type of
\emph{quantum deficit}, being the minimum cost, measured in bits of
entropy production, required by a decoupling process consisting of
local unitaries and local orthogonal measurements only.

A second, quite different, definition of decoupling task is adopted in
Ref.~\cite{quit}: here the input state $\r$ is drawn from a
non-trivial set of possible input states (constructed as the orbit of
different seed states under the action of a unitary representation of
a group), while the decoupling map can act \emph{globally} and is
required to achieve exact decoupling \emph{always} (there is no
approximation here). It is then shown that some seed states induce
orbits for which the optimal decoupling map cannot be different from
the completely depolarizing channel which transforms every state into
the maximally mixed one. Correspondingly, such correlations are called
\emph{unerasable}, as opposed to \emph{erasable} correlations, which
are those that can be decoupled by a non-trivial map. Such a
distinction hence arises as an algebraic/geometrical property of the set of quantum states analyzed, depending on the seed state and the group acting on it.

The decoupling task we propose here is independent of both previously mentioned approaches. In fact, as anticipated in the Introduction, when performing PQD, we monitor not only the decrease of correlations between $R$ and $A$, but also the corresponding \emph{increase} of correlations between $R$ and the environment (causing the decoupling \footnote{It should be clear that non-trivial decoupling processes are possible only in presence of an environment interacting with the system. A closed evolution, in fact, does not induce any change in the correlations.}). In fact, it is not difficult to see that the group-averaging channel $A\to B$ of the form
\begin{equation}
  \mathcal{R}:\sigma\mapsto\int_{\mathbb{SU}(d)}\d g\ U_g\sigma U_g^\dag=\frac{\openone}d,\qquad\forall\sigma,
\end{equation}
of which Eq.~(\ref{eq:e-rand}) aims at being a faithful approximation,
when applied on $A$, does not \emph{truly destroy} correlations
between $A$ and $R$, but simply \emph{transfers} them to the
environment, so that $R$ is no more correlated with $A$, but is
correlated with the environment. This can be explicitly seen by
writing the Stinespring purification~\cite{stine} of the channel $\mathcal{R}_A$
as an (essentially uniquely defined) isometry $V_A:A\to BE_1E_2$,
with $V_A^\dag V_A=\openone_A$, acting as
\begin{equation}
V_A|\psi_A\>=|\Psi^+_{BE_1}\>\otimes|\psi_{E_2}\>,\qquad\forall\psi,
\end{equation}
where $\Psi^+$ is a maximally entangled state. One can check that
indeed it holds
\begin{equation}
\mathcal{R}(\sigma_A)=\Tr_{E_1E_2}[V_A\sigma_AV_A^\dag],\qquad\forall\sigma.
\end{equation}
On the other hand,
\begin{equation}
\Tr_{B}[V_A\sigma_AV_A^\dag]=\frac{\openone_{E_1}}d\otimes\sigma_{E_2},\qquad\forall\sigma.
\end{equation}
In other words, the induced channel $A\to E_2$ is \emph{noiseless}:
all correlations initially shared with $R$ are now \emph{perfectly}
transferred to the environment subsystem $E_2$. (In fact, the same
phenomenon happens with all channels preparing some fixed state
regardless of the input.) We conclude that the approach adopted in Ref.~\cite{groisman} constitutes the opposite of what we want to achieve with PQD: in defining the latter, in fact, we will impose an upper bound on the amount of correlations transferred into the environment as a consequence of the decoupling process.

At this point, the natural question arises---is the constraint we are going to impose well balanced? In other words, does it induce some non-trivial situations where we can truly eliminate correlations, without merely `jettisoning' them into the environment, and some non-trivial situations where we cannot? As a first motivating example, let us consider the extreme case where the initial shared state is a pure state $|\Psi_{RA}\>$. Again, due to Stinespring's theorem~\cite{stine}, whatever deterministic transformation Alice could engineer can always be represented by an isometry $V_A:A\to BE$, with $V_A^\dag V_A=\openone_A$, the share $B$ being Alice's output, the share $E$ being the environment. In other words, the initial bipartite pure state $|\Psi_{RA}\>$ gets transformed into the tripartite pure state \begin{equation}
  |\Phi_{RBE}\>:=(\openone_R\otimes V_A)|\Psi_{RA}\>.
\end{equation}
Thanks to the identity~\cite{haya}
\begin{equation}
  I^{X:Y}(\psi_{XYZ})+I^{X:Z}(\psi_{XYZ})=2S^X(\psi_{XYZ}),
\end{equation}
where $\psi_{XYZ}$ is a tripartite pure state, we see that
\begin{eqnarray}\label{eq:conservation}
  I^{R:B}(\Phi_{RBE})+I^{R:E}(\Phi_{RBE})&=2S^R(\Phi_{RBE})\\
  &=I^{R:A}(\Psi_{RA}),
\end{eqnarray}
that shows that, if the initial state is pure, total correlations
cannot be eliminated but only moved from one system to another: pure
state correlations constitute a \emph{conserved quantity}.

As a second motivating example, for which correlations are instead
truly eliminable, let us consider the classically $A\to B$ correlated
state~\cite{malloppo}
\begin{equation}\label{eq:abcc1}
  \r=\sum_ip_i\varrho_R^i\otimes|i\>\<i|_A,
\end{equation}
where vectors $|i\>$'s are orthonormal. Then, by applying on $A$ the
Stinespring's isometry
\begin{equation}\label{eq:CC-cancelling}
  M_A:=\sum_i|e_B^i\>\<e_A^i|\otimes|e_E^i\>,
\end{equation}
where $B\cong E\cong A$ ($\dim\sH_A=d$) and the vectors $|e^i\>$'s are orthonormal vectors such that $|\<e^i|j\>|^2=1/d$ for all $i,j$, the resulting
tripartite output state
\begin{equation}
  \s:=(\openone_R\otimes M_A)\r(\openone_R\otimes M_A^\dag)
\end{equation}
satisfies
\begin{equation}
\varsigma_{RB(E)}=\varrho_R\otimes\frac{\openone_{B(E)}}{d}.
\end{equation}
In other words, by applying $M_A$, we obtained perfect decoupling
between $B$ and $R$ (since the state $\varsigma_{RB}$ is
factorized), however without correlating $R$ with $E$ (since also the state $\varsigma_{RE}$ is
factorized). In this case, we say that the initial
correlations in $\r$ have been \emph{perfectly eliminated}.

In the rest of the paper we will study what happens when initial correlations belong to a general bipartite quantum state. The general situation will turn out to lie in between the two preceding examples, in the sense that we will be able to divide total correlations into those which are \emph{eliminable} and those which are \emph{ineliminable}, the latter representing some sort of conserved quantity behaving like the correlations present in pure bipartite states.

\section{Private quantum decoupling: mathematical definition}\label{sec:3}

\subsection{Ineliminable correlations}

According to the fact that in our scenario $R$ acts as the passive reference system with respect to which information is measured, the goal of approximate PQD is to minimize correlations with $R$ by acting with a channel (i.~e. with a completely positive, trace-preserving map) only on subsystem $A$ of $\varrho_{RA}$, in such a way that the amount of correlations transferred to the environment during the decoupling process is upper bounded by a privacy parameter $\eps\ge 0$. We call this task $\eps$-PQD. The appropriate tool to analyze such a task is provided by the Stinespring dilation of a channel~\cite{stine}: given an input Hilbert space $\sH_A$ (of finite dimension $d_A$), a channel $\mD$, and an output Hilbert space $\sH_B$ (of finite dimension $d_B$), there always exists a finite dimensional auxiliary Hilbert space $\sH_E$ (of dimension $d_E\le d_Ad_B$) and an isometry $V_\mD:\sH_A\mapsto\sH_B\otimes\sH_E$, $V_\mD^\dag V_\mD=\openone_A$, such that $\mD(\sigma)=\Tr_E[V_\mD\sigma V_\mD^\dag]$, for all states $\sigma$ on $\sH_A$. Such an isometric extension $V_\mD$ is unique up to local unitary transformations on $\sH_E$, which represents the environment interacting with the system during the open evolution described by $\mD$. Before proceeding we shall notice that since everything here is finite-dimensional, minima and maxima appearing in the following are all achievable.

In practise, $A$ applies the isometry $V_\mD$ on her share, keeps the $B$ part and discards $E$. Accordingly, $\eps$-PQD is mathematically characterized by the following quantity:
\begin{equation}\label{eq:definition}
  \Xi_A(\varrho_{RA};\eps):=\min_{V_A\in\mathcal{V}_\eps(\varrho_{RA})}I^{R:B}(\s),
\end{equation}
where $\s:=(\openone_R\otimes V_A)\varrho_{RA}(\openone_R\otimes V_A^\dag)$, and $\mathcal{V}_\eps(\varrho_{RA})$ is the set of isometries $V_A:\sH_A\mapsto\sH_B\otimes\sH_E$ such that $I^{R:E}(\s)\le\eps$ and, without loss of generality since the roles of output subsystems $B$ and $E$ in the definition~(\ref{eq:definition}) can be exchanged, $I^{R:E}(\s)\le I^{R:B}(\s)$ In formula,
\begin{eqnarray}
\mathcal{V}_\eps(\varrho_{RA}):=&\bigg\{V_A:\sH_A\mapsto\sH_B\otimes\sH_E\bigg|&I^{R:E}(\s)\le\eps\nonumber\\
&&\&\ I^{R:E}(\s)\le I^{R:B}(\s)
\bigg\}.\nonumber
\end{eqnarray}
The non-negative quantity $\Xi_A(\varrho_{RA};\eps)$, which we call $\eps$-\emph{ineliminable information} (or, equivalently, $\eps$-\emph{ineliminable correlations}), measures the amount of correlations with the Referee that Alice cannot eliminate, without discharging into the environment more than $\eps$ bits of them. We will refer to the parameter $\eps$ as the \emph{privacy level parameter}: the two extreme cases, that is, $\eps=0$ and $\eps=\infty$, correspond to \emph{perfect} PQD and to \emph{advantage preserving} PQD, respectively. In the following, for sake of clarity, we will denote $\Xi_A(\varrho_{RA};\infty)$ simply as $\Xi_A(\varrho_{RA})$. Notice that $\Xi_A(\varrho_{RA};\eps)=0$ if and only if $\Xi_A(\varrho_{RA})=0$.

\subsection{Few properties at a glance}

Already from the definition~(\ref{eq:definition}), we can see that
$\Xi_{A}(\varrho_{RA};\eps)$ is invariant under local unitary
operations, that is
\begin{equation}\label{eq:uni-inv}
  \Xi_{A}\left((U_R\otimes W_A)\varrho_{RA}(U_R^\dag\otimes W_A^\dag);\eps\right)=\Xi_{A}(\varrho_{RA};\eps).
\end{equation}
Moreover, for all $\eps_1\ge\eps_2$,
\begin{equation}\label{eq:simple-upper-bound}
  0\le\Xi_{A}(\varrho_{RA};\eps_1)\le\Xi_{A}(\varrho_{RA};\eps_2)\le I^{R:A}(\varrho_{RA}),
\end{equation}
since all the values of $I^{R:B}(\s)$ that are achievable with the
constraint $I^{R:E}(\s)\le \eps_2$ are also achievable
with the looser constraint $I^{R:E}(\s)\le \eps_1$, but
not viceversa, while the upper bound $I^{R:A}(\varrho_{RA})$ is trivially
achieved when Alice does nothing at all, in such a way that
$I^{R:E}(\varsigma_{RBE})=0$.

\subsection{Private local randomness is a
  resource}\label{sec:4a}

Implicitly, by giving Alice the possibility of performing local
isometric embeddings, we are providing her with free access to local
pure states: indeed, an isometric embedding is nothing but a unitary
interaction of the system with some pure ancillary state.

On the contrary, let us think for a while to the opposite situation, like the one considered in Ref.~\cite{malloppo,local_info} in the context of local purity distillation, where Alice is granted unlimited access to local randomness, that is, she can freely create maximally mixed states. In particular, let us consider such a local randomness as being \emph{private}, i.~e. factorized from all other parties taking part---either actively or passively---into the protocol (including the adversary Eve). Within this alternative scenario, suppose that Alice and the Referee initially share some bipartite two-qubits state $\varrho_{RA}$. Since we allow local private randomness for free, we can actually consider the state \begin{equation}
  \varrho_{RA\tilde A}:=\r\otimes\frac{\openone_{\tilde A}}4,
\end{equation}
where $\tilde A\cong\mathbb{C}^4$ belongs to Alice, namely, Alice
happens to be provided with two extra-bits of private randomness. The
idea is now simple: Alice can use these two extra-bits in order to
securely decouple $A$ from $R$. The decoupling isometry $V_{A\tilde
  A}:A\tilde A\to A\tilde A$ Alice has to perform is given by
\begin{equation}
  V_{A\tilde A}:=\sum_{i=0}^3\sigma^i_A\otimes|i\>\<i|_{\tilde A},
\end{equation}
where the $\{\sigma^i\}_i$ correspond to the Pauli's matrices
$\{\openone,\sigma^x,\sigma^y,\sigma^z\}$. Written $\varsigma_{RA\tilde
  A}:=(\openone_R\otimes V_{A\tilde A})\varrho_{RA\tilde A}(\openone_R\otimes V_{A\tilde
  A}^\dag)$, it is easy to check that
\begin{equation}
  \varsigma_{R\tilde A}=\varrho_R\otimes\frac{\openone_{\tilde A}}4\qquad\&\qquad\varsigma_{RA}=\varrho_R\otimes\frac{\openone_A}2,
\end{equation}
that implies $\Xi_{A\tilde A}(\varrho_{RA\tilde A};\eps)=0$, for all $\eps$. Then, two extra-bits of private randomness in $\tilde A$ suffice to securely decouple any two-qubit state $\varrho_{RA}$ shared between $R$ and $A$, no matter how correlated is the state $\varrho_{RA}$. Notice that this is in agreement with Ref.~\cite{groisman}. It is important at this point to stress that, in order to eliminate correlations, the randomness in $\tilde A$ has to loose its privacy, since in general $\tilde A$ has to get correlated with $A$---in fact, $\varsigma_{A\tilde
  A}=(1/4)\sum_i\sigma^i_A\rho_A\sigma^i_A\otimes|i\>\<i|_{\tilde A}$. This means that, even if the reduced
state of $\tilde A$ still looks maximally mixed after the decoupling
process, it does not represent anymore a fresh source of
\emph{private} randomness: in other words, there is no
\emph{catalysis} occurring here.

We can conclude saying that, if private local randomness is provided for free, correlations can always be perfectly eliminated: in this sense, the framework of PQD implicitly assumes that local private randomness has to be considered as a \emph{resource}.

One final remark: from the preceding example, one should not jump to
the conclusion that, in order to perfectly eliminate $n$ bits of total
correlations, at least $n$ bits of extra-randomness are \emph{always}
needed. It is indeed possible to securely decouple a maximally pure
state of two qubits, hence carrying \emph{two bits} of total correlations,
using only \emph{one bit} of extra randomness. See Subsection~\ref{sec:economy}
for details.

\section{General bounds and a monogamy relation}\label{sec:4}

We now face the general problem of quantifying the amount of ineliminable correlations present in a given state $\varrho_{RA}$ which is neither pure nor simply classically correlated. The following Proposition exhibits a useful lower bound on ineliminable correlations:

\begin{prop}\label{prop:1}
  For any given state $\varrho_{RA}$, it holds that
  \begin{equation}\label{eq:statement1}
   \Xi_A(\varrho_{RA};\eps)\ge\max\{2I_c^{A\to R}(\varrho_{RA})-\eps,I_c^{A\to R}(\varrho_{RA}) ,0\},
  \end{equation}
and
 \begin{equation}\label{eq:statement2}
   \Xi_A(\varrho_{RA})\ge\max\{I_c^{A\to R}(\varrho_{RA}),0\}.\ \square
 \end{equation}
\end{prop}
{\bf Proof.} We introduce a purification $|\Psi_{SRA}\>$ of
$\varrho_{RA}$. Then, any isometry $V_A:\sH_A\mapsto\sH_B\otimes\sH_E$
produces a four-partite pure state
$|\Upsilon_{SRBE}\>:=(\openone_{SR}\otimes V_A)|\Psi_{SRA}\>$. Since
both $\Psi_{SRA}$ and $\Upsilon_{SRBE}$ are purifications of the same
state $\varrho_{R}$, it is easy to check (by direct inspection) that:
\begin{equation}\label{eq:equaz}
  I(R\!:\!A)+
  I(R\!:\!S)= I(R\!:\!B)+ I(R\!:\!SE).
\end{equation}
Notice that the notation $I(R\!:\!S)$ is not ambiguous since
$I^{R:S}(\Upsilon_{SRBE})=I^{R:S}(\Psi_{SRA})$. Plugging
into~(\ref{eq:equaz}) the chain rule $I(R\!:\!SE)=I(R\!:\!S)+I(R\!:\!E|S)$, we
obtain $I(R\!:\!B)=I(R\!:\!A)-I(R\!:\!E|S)$. A second application of
the chain rule leads to the estimate $I(R\!:\!E|S)=I(E\!:\!R|S)\le
I(E\!:\!RS)=S(SR)+S(E)-S(B)=S(A)+S(E)-S(B)$, hence to
\begin{equation}\label{eq:equaz2}
  I(R\!:\!B)\ge I_c^{A\to R}(\varrho_{RA})+S(B)-S(E).
\end{equation}
Along exactly the same lines, we also obtain
\begin{equation}\label{eq:equaz3}
  I(R\!:\!E)\ge I_c^{A\to R}(\varrho_{RA})+S(E)-S(B).
\end{equation}
The case of $\eps$-PQD requires $I(R\!:\!E)\le\eps$ and $I(R\!:\!E)\le I(R\!:\!B)$. The first condition, plugged into~(\ref{eq:equaz3}), gives $S(B)-S(E)\ge I_c^{A\to R}(\varrho_{RA})-\eps$, which in turns, plugged into~(\ref{eq:equaz2}), gives $ I(R\!:\!B)\ge 2I_c^{A\to R}(\varrho_{RA})-\eps$. The second condition, together with Eqs.~(\ref{eq:equaz2}) and~(\ref{eq:equaz3}), gives $I(R\!:\!B)\ge I_c^{A\to R}(\varrho_{RA})$. This proves Eq.~(\ref{eq:statement1}). The case of advantage preserving PQD, on the other hand, only requires the second condition $I(R\!:\!E)\le I(R\!:\!B)$, which proves Eq.~(\ref{eq:statement2}).\ $\blacksquare$\medskip

The following Proposition refines the upper bound~(\ref{eq:simple-upper-bound}):

\begin{prop}\label{prop:2}
For any given state $\varrho_{RA}$, it holds that
\begin{equation}
\Xi_{A}(\varrho_{RA})\le\frac{I^{R:A}(\varrho_{RA})}{2}.\ \square
\end{equation}
\end{prop}
{\bf Proof.}
Let us consider isometries of the form $V_{\povm{P}}:=\sum_m(|m_B\>\otimes|m_E\>)\<\phi^m_A|$, where the (in general neither normalized nor orthogonal) vectors $\{|\phi^m_A\>\}_m$ form a rank-one POVM (positive operator valued measure), i.~e. $\sum_m\phi^m_A=\openone_A$, while the vectors $\{|m\>\}_m$ are orthonormal. Isometries of such a form give $I(R\!:\!B)=I(R\!:\!E)$ by construction. This means that the application of an isometry like $V_\povm{P}$ automatically constitutes a suitable candidate for advantage preserving PQD. This means that $\Xi_A(\varrho_{RA})\le\min_{\povm{P}}I(R\!:\!B)$, where the minimum is taken over all rank-one POVM $\povm{P}$. To conclude the proof, we just notice that, by its very definition, the quantity $\min_{\povm{P}}I(R\!:\!B)$ turns out to be equal to the so-called \emph{unlocalizable entanglement} $E_u^\leftarrow(\r)$ defined in Ref.~\cite{polygamy}, where it is also proved to satisfy $E_u^\leftarrow(\r) \le I(R\!:\!A)/2$. $\blacksquare$\medskip

Due to Proposition~\ref{prop:2}, we discover the following:
\begin{corollary}
  According to definition~(\ref{eq:definition}), it is unnecessary to consider values for the privacy parameter $\eps>I^{R:A}(\r)/2$. $\square$
\end{corollary}

Another interesting consequence of Proposition~\ref{prop:2} is the following:

\begin{corollary}[Monogamy relation]\label{coro:monog}
For any given tripartite pure state $|\Psi_{RAB}\>$, for $\rho_{RA(B)}:=\Tr_{B(A)}[\Psi_{RAB}]$, it holds that
\begin{equation}\label{eq:monogamy}
\Xi_A(\rho_{RA})+\Xi_B(\rho_{RB})\le S(R),
\end{equation}
namely, ineliminable correlations are monogamous. $\square$
\end{corollary}

\noindent{\bf Proof.} Due to Proposition~\ref{prop:2}, $\Xi_A(\rho_{RA})\le I(R\!:\!A)/2$ and $\Xi_B(\rho_{RB})\le I(R\!:\!B)/2$. Simply by the definition of quantum mutual information then, we obtain
\begin{eqnarray}
&\Xi_A(\rho_{RA})+\Xi_B(\rho_{RB})\nonumber\\
\le&\frac 12\left[S(R)+S(A)-S(RA)+S(R)+S(B)-S(RB)\right]\nonumber\\
=&\frac 12\left[S(R)+S(A)-S(B)+S(R)+S(B)-S(A)\right]\nonumber\\
=&S(R),\nonumber
\end{eqnarray}
where we have made use of the fact that the joint state $|\Psi_{RAB}\>$ is pure, so that $S(RA)=S(B)$ and $S(RB)=S(A)$. $\blacksquare$\medskip

A relation like that in~(\ref{eq:monogamy}) is usually referred to as a \emph{monogamy} relation, in that it implies that ineliminable correlations cannot be shared freely among the parties of a multipartite state. Such a property, highly desirable for candidate measures of `quantum' correlations, is a distinctively quantum feature---indeed, classical correlations can be freely distributed. In Section~\ref{sec:7} we will discuss in more detail about the possibility, encouraged by Corollary~\ref{coro:monog}, of considering ineliminable correlations as a measure of the `quantumness' of the correlations present in an arbitrary bipartite quantum state.

\section{Asymptotic scenario and achievable rates region}\label{sec:5}

As it often happens, the analysis of asymptotic scenarios is considerably easier than its finite counterpart. In particular, when infinitely many i.i.d. copies of the same resource are provided, a wealth of exact results are known in the literature. In this case, we define the $\eps$-\emph{ineliminable information rate} by regularizing definition~(\ref{eq:definition}) as follows:
\begin{equation}\nonumber
  \Xi_A^\infty(\varrho_{RA};\eps):=\lim_{n\to\infty}\frac 1n\min_{V_{A_n}\in\mathcal{V}_{\eps_n}(\varrho^{\otimes n}_{R_nA_n})}I^{R_n:B_n}(\sn),
\end{equation}
where $\sn:=(\openone_{R}^{\otimes n}\otimes V_{A_n})\varrho^{\otimes n}_{R_nA_n}(\openone_{R}^{\otimes n}\otimes V_{A_n}^\dag)$, $\mathcal{V}_{\eps_n}(\varrho^{\otimes n}_{R_nA_n})$ is the set of isometries $V_{A_n}:\sH_A^{\otimes n}\mapsto\sH_{B_n}\otimes\sH_{E_n}$ such that $I^{R_n:E_n}(\sn)\le\eps_n$ and (without loss of generality, as noticed before) $I^{R_n:E_n}(\sn)\le I^{R_n:B_n}(\sn)$, and $\eps:=\lim_{n\to\infty}\eps_n/n$. According to this definition, $\eps$ becomes the \emph{privacy rate}: the case where only the constraint $I^{R_n:E_n}(\sn)\le I^{R_n:B_n}(\sn)$ is considered will be denoted as $\Xi_A^\infty(\varrho_{RA})$. Again, $\Xi_A^\infty(\varrho_{RA};\eps)=0$ if and only if $\Xi_A^\infty(\varrho_{RA})=0$. In what follows, we will characterize the achievable rates region for PQD, that is, the set of allowed pairs $\left(\Xi_A^\infty(\varrho_{RA};\eps),\eps\right)$, for any given initial state $\varrho_{RA}$.

A first characterization is given by Proposition~\ref{prop:1}, which, due to the extensivity of QMI and coherent information, i.~e. $I^{R_n:A_n}(\varrho^{\otimes n}_{R_nA_n})=nI^{R:A}(\varrho_{RA})$ and $I_c^{A_n\to R_n}(\varrho^{\otimes n}_{R_nA_n})=nI_c^{A\to R}(\varrho_{RA})$, implies that
 \begin{equation}\nonumber
   \Xi_A^\infty(\varrho_{RA};\eps)\ge\max\{2I_c^{A\to R}(\varrho_{RA})-\eps,I_c^{A\to R}(\varrho_{RA}),0\},
  \end{equation}
and
 \begin{equation}\label{eq:becomes-strict}
   \Xi_A^\infty(\varrho_{RA})\ge\max\{I_c^{A\to R}(\varrho_{RA}),0\}.
 \end{equation}
In fact, the bound~(\ref{eq:becomes-strict}) can be proved to be an equality:
\begin{prop}\label{prop:ap}
For any given state $\rho_{RA}$, it holds that
\begin{equation}\nonumber
  \Xi^\infty_A(\varrho_{RA})=\max\{I_c^{A\to R}(\varrho_{RA}),0\}.\ \square
\end{equation}
\end{prop} {\bf Proof.} We continue here from where the proof of Proposition~\ref{prop:2} ended. There, we showed that $\Xi_A(\varrho_{RA})\le\min_{\povm{P}}I(R\!:\!B)$, where the minimum is taken over all rank-one POVM $\povm{P}$. The quantity $\min_{\povm{P}}I(R\!:\!B)$ is itself a well-defined quantity with its own well-defined regularized version, which we denote by $\widetilde\Xi_A^\infty(\varrho_{RA})\ge \Xi_A^\infty(\varrho_{RA})$. Let us now introduce a purification $|\Psi_{SRA}\>$ of $\varrho_{RA}$ and define $\varrho_{SR}:=\Tr_A[\Psi_{SRA}]$. As shown in Ref.~\cite{kw}, the monogamy formula holds $\widetilde\Xi_A^\infty(\varrho_{RA})+E_A^\infty(\varrho_{SR})=S(R)$, where $E_A^\infty(\varrho_{SR})$ is the so-called \emph{rate of
  entanglement of assistance}, which has been proved in Ref.~\cite{ent-ass} to be equal to $\min\{S(S),S(R)\}$. Since $S(S)=S(RA)$, we have $\widetilde\Xi_A^\infty(\varrho_{RA})=\max\{I_c^{A\to
  R}(\varrho_{RA}),0\}$, and, due to Eq.~(\ref{eq:becomes-strict}), we obtain the statement of the
proposition. $\blacksquare$\medskip

\begin{figure}[h]
\begin{center}
\includegraphics[width=8cm]{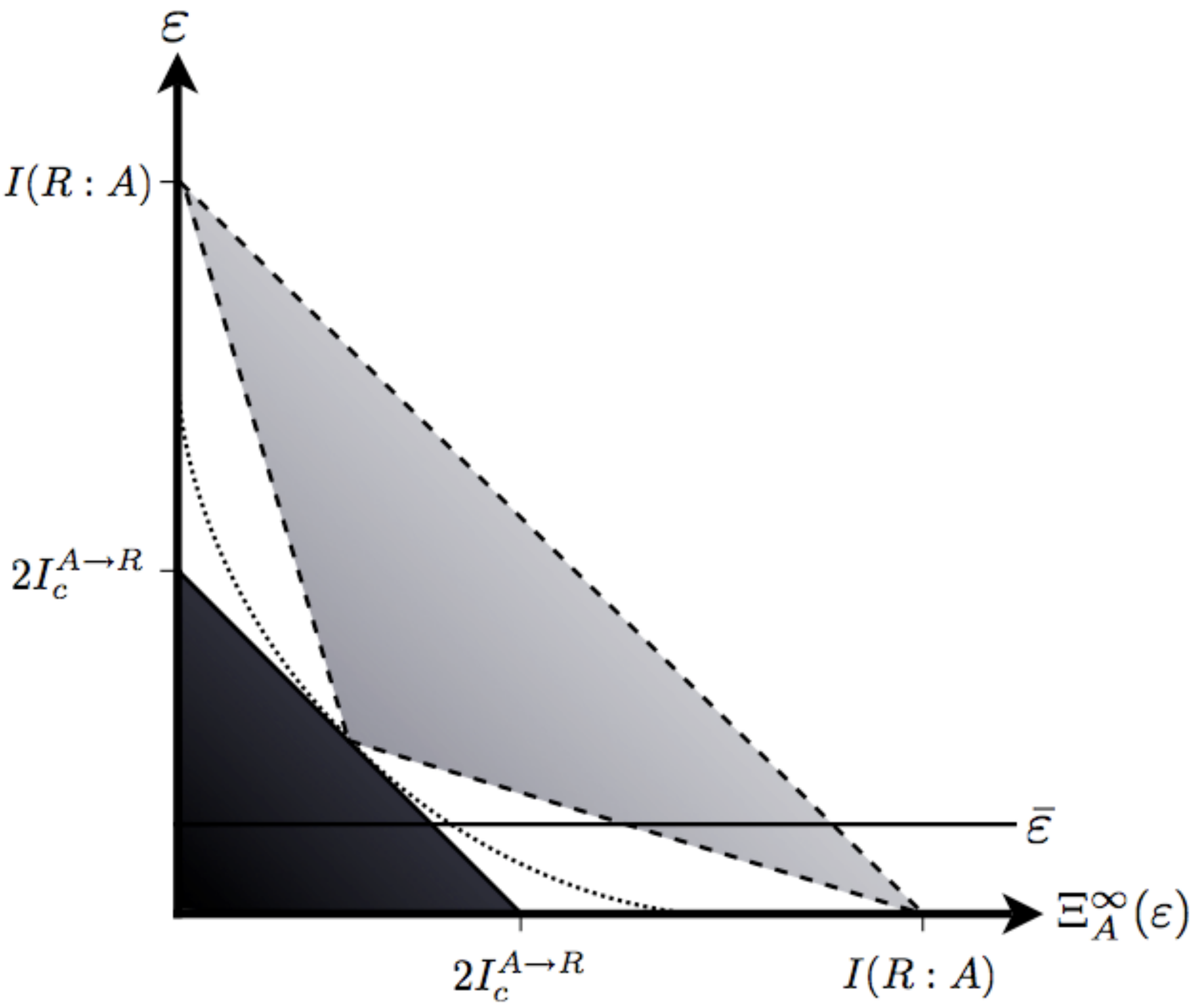}
\caption{The figure depicts the asymptotically achievable rates region for PQD for a given $\varrho_{RA}$, that is, the set of allowed pairs $\left(\Xi_A^\infty(\varrho_{RA};\eps),\eps\right)$, for any given initial state $\varrho_{RA}$. The symmetry about the bisector reflects the possibility of exchanging the roles of subsystems $B$ and $E$ in Eq.~(\ref{eq:definition}). The dark-grey shaded area around the origin corresponds to rates which are forbidden due to Proposition~\ref{prop:1}. The light-grey shaded area, instead, corresponds to rates achievable via the result in Proposition~\ref{prop:ap} and time-sharing. The white areas in between are not characterized yet, as well as the boundary of the achievable rates region, which could well be given by some curve similar to the dotted one. When the privacy rate is set to some $\bar\eps<\infty$, the corresponding achievable rates region is further constrained to lie below the line $\eps=\bar\eps$. Finally, note that, for pure states, $I(R\!:\!A)=2I_c^{R\to A}$, namely, the achievable rates region collapses onto the most external dashed line, according to the fact that correlations carried by bipartite pure states constitute a conserved quantity. On the other hand, for separable states, $I_c^{R\to A}\le 0$, namely, the achievable rates region fills the octant: in this case, perfect PQD, i.~e. the origin $(0,0)$, is achievable (see Corollary~\ref{coro:sep}).}
\label{default}
\end{center}
\end{figure}

Our analysis hence led us to a situation like the one depicted in Figure~\ref{default}, where the achievable rates region for a given initial state $\varrho_{RA}$ is sketched.

As we noticed before, the condition $\Xi_A^\infty(\varrho_{RA})=0$ is equivalent to saying that $A$ can be privately decoupled from $R$, in the limit of infinitely many copies provided. On the other hand, it is known that, for any separable state $\varrho_{RA}$, $\max\{I_c^{R\to A},I_c^{A\to R}\}\le 0$,~\cite{horo}. This provides an intriguing connection between PQD and entanglement theory, stated in the following corollary of Proposition~\ref{prop:ap}:
\begin{corollary}[Ineliminable versus separable correlations]\label{coro:sep}
  For any separable state $\varrho_{RA}$, $\Xi_A^\infty(\varrho_{RA})=0$. In other words, the presence of
  asymptotically ineliminable information is a signature of quantum
  entanglement. $\square$
\end{corollary}

\section{Some examples}\label{sec:6}

The following examples are presented in order to show that the notion
of ineliminable correlations cannot be straightforwardly explained in
terms of entanglement only, as soon as one leaves the pure state
case. It is however very hard to find explicit counterexamples, as the
minimization in Eq.~(\ref{eq:definition}) is difficult to be explicitly
solved in general.

\subsection{Pure states}

As we already noticed, Propositions~\ref{prop:1}-\ref{prop:ap} imply that, for any pure bipartite
state $|\Psi_{RA}\>$, it holds
\begin{equation}
\Xi_{A}(\Psi_{RA})=\Xi^\infty_{A}(\Psi_{RA})=S(R),
\end{equation}
namely, ineliminable correlations equal the entropy of entanglement.

\subsection{Classically correlated states}\label{sec:class_corr}

When $A$ possesses purely classical information about $R$, i.~e. when the shared state $\r$ is classically $A\to R$ correlated, that is~\cite{malloppo,gros-mor} \begin{equation}
\label{eq:abcc}
\r=\sum_ip_i\varrho_R^i\otimes|i\>\<i|_A,
\end{equation}
for orthonormal $|i\>$'s, we already saw at the end of Section~\ref{sec:2} that \begin{equation}
  \Xi_{A}(\r)=\Xi_{A}(\r;\eps)=0,
\end{equation}
that is, classical correlations can be perfectly shredded.

\subsection{Random-unitary channels}\label{subsec:randu}

Let us consider a quantum system $A$, whose state is initially described be the density matrix $\sigma$, undergoing a channel $\mathcal{T}:A\to B$. Let $|\Psi_{RA}\>$ be the purification of $\sigma_A$, where the system $R$ plays the role of a reference that does not change in time. Moreover, let $W_A:A\to BE$ be the Stinespring's isometry~\cite{stine} purifying the channel $\mathcal{T}$, that is
\begin{equation}
  \mathcal{T}_A(\sigma_A)=\Tr_{E}[W_A\sigma_A W_A],\qquad\forall\sigma.
\end{equation}
Let us denote as
$|\Phi_{RBE}\>:=(\openone_R\otimes W_A)|\Psi_{RA}\>$ the tripartite
pure state finally shared among the output $B$, the environment
$E$, and the reference $R$.

If the channel $\mathcal{T}$ is a closed evolution, namely, if it is described
by one isometry only, then the reference $R$ is completely decoupled
from the environment $E$. Let us suppose now that the only error
occurring in the whole process is due to a classical shuffling,
resulting in a classical randomization of different possible
isometries: then, the resulting evolution will not be described by one
particular isometry, as in the closed evolution case, but rather by a
mixture of such isometries. This kind of noisy evolutions are called
\emph{random-unitary channels} and act like
\begin{equation}\label{eq:randunit}
\mathcal{R}:\sigma\mapsto\sum_ip_iV_i\sigma V_i^\dag,
\end{equation}
where $p_i$ is a probability distribution and $V_i:A\to B$ are
isometries. The following questions arise naturally: which kind of
correlations between the reference and the environment cause (or,
depending on the point of view, are caused by) such a `classical'
error? Which properties do these correlations satisfy? Can we ascribe
a `classical character' to these correlations?

It is known~\cite{erasure-corr} that a channel $\mathcal{R}:A\to B$ admits (on
the support of the input state $\sigma$) a random-unitary Kraus
representation as in Eq.~(\ref{eq:randunit}) if and only if there
exists a rank-one POVM $\povm{P}_E:=\{\varphi_E^m\}_m$, $\sum_m\varphi_E^m=\openone_E$, on the environment $E$ such
that
\begin{equation}
  \varrho_R^m\propto\varrho_R,\qquad\forall m,
\end{equation}
where $\varrho_R^m:=\Tr_{BE}[\Phi_{RBE}\ (\openone_{RB}\otimes\varphi_E^m)]$. This implies that
\begin{equation}\label{eq:implic1}
  \mathcal{R}\textrm{ random-unitary}\ \Rightarrow\ \Xi^{A_2}(\tau^{A_2B};\eps)=0,\qquad\forall\eps\ge 0.
\end{equation}
In other words, random-unitary channels create, between the
reference system $R$ and the environment $E$, correlations that are
perfectly eliminable with a local action on $E$ only.

  It is important now to stress that, in general, the form of the reference-environment joint state $\Tr_B[\Phi_{RBE}]$, originating from the purification of a random-unitary channel $\mathcal{R}:A\to B$, can in principle be different from that of a classically $E\to R$ correlated state as in Eq.~(\ref{eq:abcc}). In other words, random-unitary channels induce a class of states for which perfect PQD is possible that is in principle \emph{larger} than the class of classically correlated states.

\medskip\noindent {\bf Open problem.} Since the implication in~(\ref{eq:implic1}) is in one direction only, it would be interesting to characterize the class of channels inducing only perfectly eliminable correlations between the reference and the environment: is such a class strictly larger than the class of random-unitary channels? If so, does it admit an easier characterization \footnote{In fact, there is no known constructive algorithm to check whether a given channel is random-unitary or not. Only the necessary condition of being unital can be easily checked, but it is also sufficient only in the simple case of qubit channels.}?

\subsection{Mixed entangled states}\label{sec:economy}

We already saw, in Subsection~\ref{sec:4a}, that there exist entangled mixed states that can be securely decoupled for all $\eps\ge 0$. There, we needed two extra-bits of private randomness in order to securely decouple whatever two-qubits state, in agreement with the fact that a two-qubits state contains at most two bits of total correlations. However, before rushing to the conclusion that we \emph{always} need at least $n$ extra-bits of randomness to perfectly eliminate $n$ bits of total correlations, we should consider the following example where \emph{just one} extra-bit of randomness is required to securely decouple a maximally entangled pure state of two qubits (hence carrying two bits of total correlations).

Let us consider indeed the state acting on
$\mathbb{C}^2\otimes\mathbb{C}^2\otimes\mathbb{C}^2$ defined as
\begin{equation}\label{eq:gilad}
  \sigma_{RA\tilde A}:=\Psi^+_{RA}\otimes\frac{\openone_{\tilde A}}{2},
\end{equation}
where $|\Psi^+_{RA}\>:=2^{-1/2}(|00\>+|11\>)$.
For this state, even if $\tilde A$ carries only \emph{one} bit of
extra-private randomness, one can show that $\Xi_{A\tilde A}(\sigma_{RA\tilde A})=0$. The proof is easy, as the null value is
achieved by the isometry $M_{A\tilde A}:A\tilde A\to BE$, with $B\cong E\cong\mathbb{C}^4$, defined
as
\begin{equation}
  M_{A\tilde A}:=\sum_{i=1}^4|i_{B}\>\<e^i_{A\tilde A}|\otimes|i_{E}\>,
\end{equation}
where
\begin{eqnarray}\nonumber
|e^1\>&:=2^{-1/2}(|00\>+|11\>),\nonumber\\
|e^2\>&:=2^{-1/2}(|00\>-|11\>),\nonumber\\
|e^3\>&:=2^{-1/2}(|10\>+|01\>),\nonumber\\
|e^4\>&:=2^{-1/2}(|10\>-|01\>).\nonumber
\end{eqnarray}
In fact, the isometry $M_{A\tilde A}$ is coherently performing the
measurement needed to teleport the maximally mixed state $\openone/2$
from $\tilde A$ to $R$. Hence, for every outcome $i$, the reduced
state on $R$ is equal to $\openone/2$, so that, written $\s=(\openone_R\otimes M_{A\tilde A})\sigma_{RA\tilde A}(\openone_R\otimes M_{A\tilde A}^\dag)$, we get
$\varsigma^{RB}=\frac{\openone_R}2\otimes\frac{\openone_{B}}{4}$
and $\varsigma^{RE}=\frac{\openone_R}2\otimes\frac{\openone_{E}}{4}$, which yields $\Xi_{A\tilde A}(\sigma_{RA\tilde A};\eps)=0$, for all $\eps\ge0$.

\section{Discussion: ineliminable correlations, entanglement, and the `quantumness' of correlations}\label{sec:7}

Through PQD, we found a non-trivial division of total correlations, $I(A\!:\!B)$, into ineliminable ones, measured by $\Xi_{A}(\r)$, and eliminable ones, representing the rest, that is, $I(A\!:\!B)-\Xi_{A}(\r)$.

At this point, it is tempting to speculate a bit about hypothetical relations between the division of correlations into ineliminable and eliminable ones, versus the division into quantum and classical correlations~\cite{malloppo,q-c-corr,grudka,gros-mor}. We already saw how random-unitary noise, that is classical noise in the sense explained in Subsection~\ref{subsec:randu}, only induces perfectly eliminable correlations between the reference and the environment. Moreover, ineliminable correlations satisfy the two axioms required in Ref.~\cite{gros-mor} for a measure of quantumness: they are zero for classically correlated states~(\ref{eq:abcc}) and they are invariant under local unitary transformations~(\ref{eq:uni-inv}).

Also, being upper bounded by one half of the quantum mutual information and
being equal to the entropy of entanglement for pure states,
ineliminable correlations fall under the hypotheses of Theorem~2 in
Ref.~\cite{buscemi2}, which proves that, for an arbitrary bipartite
state $\r$ with $S(A)\le S(R)$, it holds
\begin{equation}\label{eq:reverse}
  0\le S(A)-I_c(A\to R)\le\mathrm{f}\left(2\sqrt{S(A)-\Xi_{A}(\r)}\right),
\end{equation}
where $\mathrm{f}(x)$ is a Fannes-type function~\cite{haya}, that is a positive, concave, continuous, monotically increasing function which depends on the dimension of the underlying Hilbert space only logarithmically and satisfies $\lim_{x\to0}\mathrm{f}(x)=0$. Eq.~(\ref{eq:reverse}) shows that, whenever the amount of ineliminable correlations in a bipartite state is sufficiently large, then such a state is necessarily entangled, since also coherent information has to be correspondingly close to the upper bound $S(A)$.

In this sense, ineliminable correlations can be considered `genuinely quantum' correlations, since Eq.~(\ref{eq:reverse}) tells us that they constitute an entanglement \emph{parameter}, in the sense explained in~\cite{ent-param}, namely, the more ineliminable correlations are present, the more the state is entangled (coherent information is the paramount example of an entanglement parameter), even though ineliminable correlations do not satisfy many natural requirements to be a proper entanglement \emph{measure}. In reinforcing this interpretation, it stands the fact, expressed by Corollary~\ref{coro:monog}, that ineliminable correlations indeed satisfy a monogamy constraint, which is another strongly distinctive feature of quantum correlations versus classical ones.

Another interesting feature of ineliminable correlations is that, for
every state $\r$, they always represent (already at the level of a single copy) \emph{at most one half} of the
total amount of correlations, that is, $\Xi_{A}(\r)\le
I(R\!:\!A)/2$ always. This fact is to be compared, once again, with what
happens for different measures of `quantum vs classical' correlations:
according to different definitions, there exist quantum states
exhibiting quantum correlations \emph{without} classical correlations,
hence representing a strikingly counterintuitive
situation~\cite{grudka,q>c}. On the contrary, assuming for a while the
definition of quantum correlations as the ineliminable ones, every
quantum state would turn out to be always more correlated classically
than quantum, hence reinforcing the common-sense intuition about
correlations.

However, in spite of this encouraging list of properties, the
existence of entangled---even maximally entangled---states with
perfectly eliminable correlations only (recall the examples analyzed
in Subsections~\ref{sec:4a} and~\ref{sec:economy}) seems to
stand as an insurmountable argument against the maybe naive statement
`what is ineliminable is quantum'. We however think that the dichotomy
proposed here can contribute to the program of understanding the
structure of total correlations, as coming from the \emph{operational}
paradigm of distant laboratories, versus the notion of entanglement,
which is the \emph{formal} property of not being separable.

\section{Conclusions}\label{sec:8}

We introduced the operational task of private quantum decoupling (PQD), which naturally arises as a model-independent description of secure disposal of information. Partial results suggest that there may be a deep connection between the theory of PQD and the theory of quantum entanglement. Further research in this direction would be in order. Moreover, it could be useful to generalize the asymptotic results presented here to the one-shot scenario, by exploiting some recently introduced tools~\cite{renner}. Also, PQD could turn out to be useful in designing quantum cryptographic protocols, where the possibility of securely erasing old data and keys is required~\cite{proactive}.

\appendix

\section*{Acknowledgments}

Stimulating discussions with N.~Datta, G.~Gour, M.~Horodecki, and J.~Oppenheim are gratefully acknowledged. This research was supported by the Program for Improvement of Research Environment for Young Researchers from Special Coordination Funds for Promoting Science and Technology (SCF) commissioned by the Ministry of Education, Culture, Sports, Science and Technology (MEXT) of Japan.

\end{document}